\RequirePackage{fix-cm}
\documentclass[smallextended]{svjour3}      
\smartqed  
\usepackage{graphicx}
\usepackage{amssymb,amsmath}
\usepackage{color}
 \journalname{General Relativity and Gravitation}
\begin{document}
\title{Gravitational radiation by point particle eccentric binary systems in the linearised characteristic formulation of general relativity 
}
\titlerunning{Gravitational Radiation From a Point particle binary system}        
\author{C. E. Cede\~no M. \and J. C. N. de Araujo}
\institute{Av. dos Astronautas, 1758, Instituto Nacional de Pesquisas Espaciais, S\~ao Jos\'e dos Campos, SP, Brazil, 12227-010 \\
 Tel.: +55-12-32086000\\
\email{eduardo.montana@inpe.br}             \\
\email{jcarlos.dearaujo@inpe.br}
}
\date{Received: date / Accepted: date}
\maketitle
\begin{abstract}
We study a binary system composed of point particles of unequal masses in eccentric orbits in the linear regime of the characteristic formulation of general relativity, generalising a previous study found in the literature in which a system of equal masses in circular orbits is considered. We also show that the boundary conditions on the time-like world tubes generated by the orbits of the particles can be extended beyond circular orbits. Concerning the power lost by the emission of gravitational waves, it is directly obtained from the Bondi's News function. It is worth stressing that our results are completely consistent, because we obtain the same result for the power derived by Peters and Mathews, in a different approach, in their seminal paper of 1963. In addition, the present study constitutes a powerful tool to construct extraction schemes in the characteristic formalism to obtain the gravitational radiation produced by binary systems during the inspiralling phase.
\keywords{General Relativity\and Characteristic formulation\and Linear Regime\and Approximation Methods}
 \PACS{04.30.-w\and , 04.20.-q}
\end{abstract}
\section{Introduction}
\label{intro}
In recent years great advances in the field of numerical relativity in the characteristic formulation have been made. In order to deal with a wide variety of interesting problems in the non-linear regime of the Einstein's field equations, such as the gravitational wave extraction algorithms \cite{RBPS10,BBSW11,BSWS11} or the gravitational collapse problem  \cite{GPW94,BD96,BD99,BPR98,BGLMW99,LFJMP01,SFP01,SFMP03,BGLMW05}, several high accurate numerical codes have been developed.
\\Despite a lack of a real meaning in the neighbourhood of the sources, the weak field approximation furnishes a good start point to create toy models that could serve to calibrate these complex and accurate codes. In addition, this approximation constitutes a powerful tool to construct extraction schemes to obtain the gravitational radiation produced by binaries in the inspiralling phase. This is done in regions that are far enough from the sources where the space-time can be essentially considered flat.  
\\The advantage of the linear regime is in the fact that the field equations become hyperbolic and hence can admit a standard variable separation, which allows to integrate them through a completely analytical way. Such a kind of procedure is well-known in the literature and allows one to find the solution for systems in which the density is given as a distribution function such as in the case of a thin shell, a point particle rotating around a Schwarzschild black-hole or two point particles orbiting each other in circular orbits due to their mutual gravitational interaction \cite{B05,BPR11,CA15}.     
\\Still concerning the use of the linear regime of the characteristic formulation in the study of binary systems, this issue has not been completely considered in the literature, since eccentricity has not been yet considered. It is worth mentioning that, as shown in the 1960s decade in the pioneer paper by Peters and Mathews \cite{PM63}, the major contribution to the power emitted by point particle binary systems is due to the eccentricity of their orbits. 
\\
In the next sections, we show how the power emitted in gravitational waves by a binary system is obtained, computed from the Bondi's News function, when eccentric orbits in point particle binary systems composed of different masses are considered.
\\In order to do so, a brief review of the characteristic formalism of null cones oriented to the future is present in Sec. II, the procedure to obtain the solution for a binary system with eccentric orbits is shown in Sec. III and the calculations of the power loss by gravitational radiation by the system is present in Sec. IV. Finally, the conclusions and other issues for further studies are presented in Sec. V.
\section{Formalism}
The characteristic formulation based on null cones oriented to the future is well-known and has been used to explore a wide variety of interesting problems (see for example  \cite{BGLW96,BPR98,BGLMW97,BGLMW99,B05,BGLMW05}). In order to avoid coordinate singularities in the angular differential operators, the {\it eth} formalism has been employed as shown in \cite{BGLW96,GPLPW97,BGLMW99,BBSW11,BPR11,BGHLW03,GBF07}. However, a quick and concise review of its more important aspects is necessary in order to fix the conventions and notations used here. The geometrised  unit system is employed, i.e., $G=c=1$. The Greek indices run from 1 to 4, labelling each coordinate, whereas the Capital Latin letters label the angular coordinates and run from 3 to 4. 
\\
The space-time is foliated into null cones emanating from a time-like central geodesic oriented to the future. The coordinates used are $x^{\mu}=(u,r,x^{A})$, where $u$ is the retarded time, which parametrises the central geodesic and labels each null cone, $r$ is the luminosity distance, which is measured along the null rays contained in these cones, and $x^{A}$ represents the angular coordinates, $(\theta,\phi)$ ($(q,p)$) for the usual spherical coordinates (for the stereographic two patches representation) and are considered as constants along the outgoing null geodesic.
Notice that the coordinates are chosen such that the hypersurfaces for $r$ and $u$ constants have an area of $4\pi r^2$.
\\In these coordinates the Bondi-Sachs metric \cite{BBM62,S62} reads,
\begin{eqnarray}
\label{bs}
ds^2&=&-\left(e^{2\beta}\left(1+\frac{w}{r}\right)-r^2h_{AB}U^AU^B\right)du^2 -2e^{2\beta}dudr 
\nonumber \\&& 
-2r^2h_{AB}U^Bdx^A du+r^2h_{AB}dx^Adx^B,
\end{eqnarray}
where $w$ and the redshift $\beta$ are related to the Newtonian gravitational potential and the square of the ADM lapse function, respectively. $U^A$ labels the shift vector between two successive null cones and the metric associated with the angular manifold (unit sphere) is represented by $h_{AB}$ ($q_{AB}$). In addition, the Bondi's gauge is imposed, i.e.  $\det(h_{AB})=\det(q_{AB})=1$; and it is required that $h_{AB}h^{BC}=q_{AB}q^{BC}=\delta_B^{~C}$. 
\\The metric of the unit sphere is decomposed as a dyadic product i.e., $q_{AB}=q_{(A}q_{B)}$, where $q_{A}$ are complex null vectors related to the tangent vectors to the unitary sphere oriented along the coordinate lines associated with the angular chart used to make the finite coverage of the sphere, i.e., $q_Aq^A=0$. Also, these vectors are chosen to satisfy $q_A\overline{q}^A=2$. Their explicit form in stereographic coordinates are
\begin{equation*}
q^A=\frac{(1+|\zeta|^2)(\delta^A_{~~3}+i\delta^A_{~~4})}{2}, \hspace{0.5cm}\zeta=\tan(\theta/2)e^{i\phi}, \hspace{0.5cm}\zeta=q+ip, 
\end{equation*}
\cite{GPLPW97,BGLMW97,BGLMW99,B05,RBLTS07,GBF07} or in spherical coordinates
\begin{equation*}
q^A=\delta^A_{~~3}+i\delta^A_{~~4} \csc\theta,
\end{equation*}
as shown in \cite{RBLTS07,HS15,BPR11}.\\
The angular part of all tensor quantities, such as the metric, the Riemann or the Ricci tensors, are projected along these tangent null vectors resulting in spin-weighted scalars. In particular, the angular metric $h_{AB}$ are decomposed in two complex $J,\overline{J}$ and a real $K$ spin-weighted scalars, i.e.
\begin{equation*}
J=q^Aq^Bh_{AB}/2, \hspace{0.5cm}\overline{J}=\overline{q}^A\overline{q}^Bh_{AB}/2, \hspace{0.5cm}K=q^A\overline{q}^Bh_{AB}/2,
\end{equation*}
and the shift vector $U^A$ is decomposed in two complex spin-weighted scalars $U$ and $\overline{U}$ as,
\begin{equation*}
U=q^AU_A, \hspace{0.5cm}\overline{U}=\overline{q}^AU_A,
\end{equation*}
where the overline indicates complex conjugation. 
\\The projection of the covariant derivative referred to $q_{AB}$ onto the vectors $q^A$ define two differential operators labelled as $\eth$ and $\overline\eth$. They can raise or lower the spin-weight of any spin-weighted scalar function $_s\Psi$ and are defined as 
\begin{equation*}
\eth\ _s\Psi=q^D\ _s\Psi_{,D}+s\Omega \ _s\Psi, \hspace{1cm}\overline{\eth}\ _s\Psi=\overline{q}^D\ _s\Psi_{,D}-s\overline{\Omega} \ _s\Psi
\end{equation*}
where $\Omega=-q^Aq^Bq_{B|A}/2$, the comma and the vertical line in the indices indicate partial and covariant differentiation associated with $q_{AB}$, respectively. For a complete and detailed review of these operators and their properties, see \cite{T07,GPLPW97,NP66,GMNRS66}. The eigenfunctions of the operator $[\eth,\overline{\eth}]$ are the spin-weighted spherical harmonics $_sY_{lm}$ defined from the usual spherical harmonics $Y_{lm}$ \cite{NP66,GMNRS66} as,
\begin{equation}
_sY_{lm}=\begin{cases}
\sqrt{\dfrac{(l-s)!}{(l+s)!}}\eth^sY_{lm} & \text{if} \hspace{0.5cm} s\ge 0 \\
(-1)^s\sqrt{\dfrac{(l-s)!}{(l+s)!}}\bar{\eth}^{-s}Y_{lm} & \text{if} \hspace{0.5cm} s< 0
\end{cases}.
\label{sYlm}
\end{equation}
However, there exists another base of eigenfunctions for the same commutator, that will result convenient because they allow to decouple completely the field equations. They are labelled as $_sZ_{lm}$ and are  constructed as linear combinations of the $_sY_{lm}$  \cite{ZGHLW03,RBLTS07,RBP13} as, 
\begin{equation}
_sZ_{lm}=\begin{cases}
\dfrac{i}{\sqrt{2}}\left((-1)^m\ _sY_{lm}+\ _sY_{l\ -m}\right) &\text{for} \hspace{0.5cm} m<0 \\
_sY_{lm}&\text{for} \hspace{0.5cm} m=0 \\
\dfrac{1}{\sqrt{2}}\left(_sY_{lm}+(-1)^m\ _sY_{l\ -m}\right) &\text{for} \hspace{0.5cm} m>0
\end{cases}.
\end{equation}
The Einstein's field equations read
\begin{equation}
E_{\mu\nu}=R_{\mu\nu}-8\pi\left(T_{\mu\nu}- g_{\mu\nu}T/2\right)=0,
\end{equation} 
and in this formalism they can be re-expressed as
\begin{subequations}
\begin{eqnarray}
&& E_{22}=0, \hspace{0.2cm}E_{2A}q^A=0, \hspace{0.2cm} \hspace{0.2cm} E_{AB}h^{AB}=0,\\ 
&& E_{AB}q^{A}q^{B}=0,\\
&& E_{11}=0, \hspace{0.2cm}E_{12}=0,\hspace{0.2cm}\hspace{0.2cm} E_{1A}q^A=0,
\end{eqnarray}
\label{f_equations}
\end{subequations}
corresponding respectively to hypersurface, evolution and constraint equations \cite{B05,RBP13,RBLTS07,BGLMW97}.
\\In stereographic-null coordinates, when the angular metric $h_{AB}$ and the shift vector are expressed in terms of the spin-weighted scalars, the Bondi-Sachs metric \eqref{bs} reads
\begin{eqnarray}
\label{bs_explicit}
ds^2&=&-\left(e^{2\beta}\left(1+\frac{w}{r}\right)-r^2(J\bar{U}^2+U^2\bar{J}+2 K U \bar{U})\right)du^2  -2e^{2\beta}dudr \nonumber\\
&&-\frac{2r^2\left((K +\bar{J}) U+(J+K) \bar{U}\right)}{1+|\zeta|^2}dq du\nonumber\\
&&-\frac{2ir^2\left((K -\bar{J}) U+(J-K)	\bar{U}\right)}{1+|\zeta|^2}dpdu+\frac{2 r^2 \left(J+2K+\bar{J}\right)}{(1+|\zeta|^2)^2}dq^2\nonumber\\
&&-\frac{4 i r^2
\left(J-\bar{J}\right)}{(1+|\zeta|^2)^2}dqdp-\frac{2 r^2 \left(J-2 K+\bar{J}\right)}{(1+|\zeta|^2)^2}dp^2.
\end{eqnarray}
\\In the weak field limit, i.e., when slight deviations from the Minkowski background $|g_{\mu\nu}|\ll|\eta_{\mu\nu}|$  are considered, and the second order terms are disregarded, the Bondi-Sachs metric is reduced to,
\begin{eqnarray}
\label{bs_lin}
ds^2&=&-\left(1-\frac{w}{r}-2\beta\right)du^2 -2(1+2\beta)dudr -2r^2\frac{(U+\overline{U})}{1+|\zeta|^2}dq du
\nonumber \\&& 
-2r^2\frac{i(U-\overline{U})}{1+|\zeta|^2}dp du+2r^2\frac{\left(2+J+\overline{J}\right)}{\left(1+|\zeta|^2\right)^2}dq^2\nonumber\\&&
-4ir^2\frac{(J-\overline{J})}{(1+|\zeta|^2)^2}dqdp-2r^2\frac{\left(-2+J+\overline{J}\right)}{\left(1+|\zeta|^2\right)^2}dp^2,
\end{eqnarray} 
which clearly can be separated as,
\begin{eqnarray}
\label{bs_lin2}
ds^2&=&-du^2-2dudr + \frac{4r^2}{\left(1+|\zeta|^2\right)^2}\left(dq^2+dp^2\right) +\left(\frac{w}{r}+2\beta\right)du^2   \nonumber\\
&&-4\beta dudr -\frac{2r^2}{1+|\zeta|^2}du\left((U+\overline{U})dq -i(U-\overline{U})dp \right)
\nonumber \\&& 
-4ir^2\frac{(J-\overline{J})}{(1+|\zeta|^2)^2}dqdp+\frac{2r^2\left(J+\overline{J}\right)}{\left(1+|\zeta|^2\right)^2}\left(dq^2-dp^2\right),
\end{eqnarray} 
showing that it corresponds to a Minkowski background plus a perturbation. 
\\
The field equations \eqref{f_equations}, corresponding to this perturbation, previously computed by Bishop in \cite{B05}, read
\begin{subequations}
\begin{align}
& 8 \pi  T_{22}=\frac{4 \beta _{,r}}{r}, \label{field_eq_1}
\\
& 8 \pi  T_{2A} q^A=\frac{\overline{\eth }J_{,r}}{2} -\eth \beta _{,r} +\frac{2 \eth \beta}{r} +\frac{\left(r^4 U_{,r}\right)_{,r}}{2r^2}, \label{field_eq_2}
\\
& 8 \pi  \left(h^{AB} T_{AB}-r^2 T\right) =-2 \eth \overline{\eth }\beta +\frac{\eth^2\overline{J} + \overline{\eth }^2 J}{2} +\frac{\left(r^4\left(\overline{\eth}U+\eth \overline{U}\right)\right)_{,r}}{2r^2}\nonumber\\
& \hspace{4.0cm}+4 \beta -2 w_{,r}, \label{field_eq_3}
\\
& 8 \pi T_{AB}  q^A q^B=-2 \eth^2 \beta + \left(r^2\eth U \right)_{,r} - \left(r^2 J_{,r}\right)_{,r} +2 r\left(rJ\right)_{,ur},\label{field_eq_4}
\end{align}
\begin{align}
& 8 \pi  \left(\frac{T}{2}+T_{11}\right)=\frac{\eth \overline{\eth }w}{2 r^3} + \frac{\eth \overline {\eth }\beta }{r^2} -\frac{\left(\eth \overline{U} + \overline{\eth} U \right)_{,u}}{2} +\frac{w_{,u}}{r^2} +\frac{w_{,rr}}{2 r} \nonumber\\
&\hspace{3.2cm} -\frac{2 \beta_{,u}}{r}+\frac{2\beta_{,r}}{r}+\beta_{,rr}-2\beta_{,ru}, \label{field_eq_5}
\\
& 8 \pi  \left(\frac{T}{2}+T_{12}\right)=
\frac{\eth \overline{\eth }\beta }{r^2} -\frac{\left(r^2\left(\eth \overline{U} + \overline{\eth   }U\right)\right)_{,r}}{4r^2}+\frac{2\beta_{,r}}{r}\nonumber\\
&\hspace{3.2cm}+\beta_{,rr}-2\beta_{,ru} +\frac{w_{,rr}}{2 r}, \label{field_eq_6}\\
& 8 \pi  T_{1A} q^A=\frac{\overline{\eth }J_{,u}}{2} -\frac{\eth^2
\overline{U} }{4} +\frac{\eth \overline{\eth }U}{4} +\frac{1}{2}\left(\frac{\eth w}{r}\right)_{,r} -\eth \beta _{,u} +\frac{\left(r^4U_{,r}\right)_{,r}}{2r^2}\nonumber\\
& \hspace{2cm}-\frac{r^2 U_{,ur}}{2} +U. \label{field_eq_7}
\end{align}
\label{field_eqs}
\end{subequations} 
\section{Binary system}
\begin{figure}
\centering
\begin{tabular}{cc}
	\includegraphics[height=4cm]{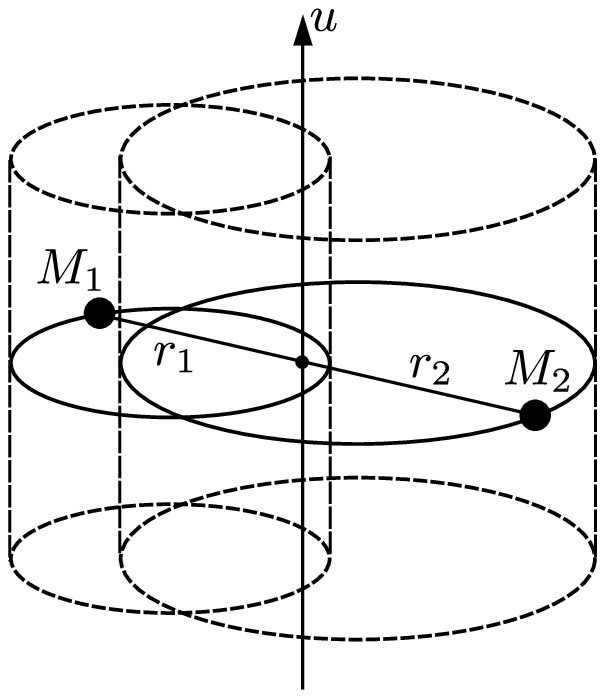}&\includegraphics[height=4cm]{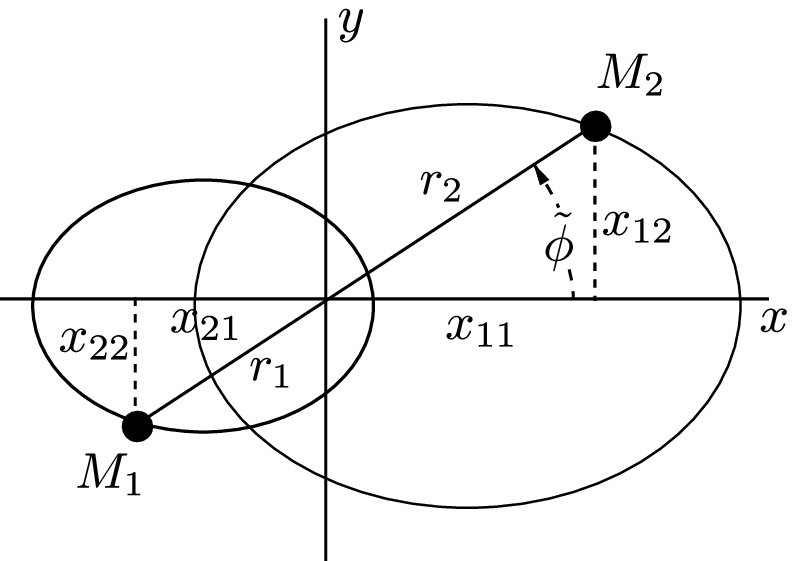}\\
	(a) & (b)
\end{tabular}
\caption{(a) Eccentric binary system with the world tubes of their orbits extended along the central time-like geodesic. (b) Top view of the point particle binary system, where the angular position $\tilde{\phi}$ is indicated.}
\label{figure1}
\end{figure}

In the weak field limit, with the background metric given by the Minkowski space-time, the conservation of the stress-energy tensor in rectangular coordinates, can be written as
\begin{equation*}
T^{\mu\nu}_{~~~;\nu}=T^{\mu\nu}_{~~~,\nu},
\end{equation*}
where the semicolon indicates covariant derivative as usual. Thus, in this regime, the gravitational field of a given system has no influence upon the motion of the matter that produces the field (as shown in standard textbooks \cite{Stephani_1990,MTW_1973}). 
Therefore, it is possible to consider particles moving along any curve in the space-time. These particles may be even producing gravitational waves. Consequently, the interaction between two particles, for example, can be  gravitational (Newtonian at first order), electromagnetic or of any kind. 
For this reason we can consider in the present study two point particles held together by their mutual gravitational interaction, moving around each other in elliptical orbits, just as sketched in figure \ref{figure1}. 
\\The masses of the binary system are responsible for the perturbation of the Minkowski space-time, generating gravitational radiation which is propagated away from the system. In order to solve the field equations for this particular situation, the time-like world tubes generated by the orbits of the point particle system must be considered. Thus, the space-time can be separated into three empty regions. The field equations are solved for such regions, and the solutions between two adjacent regions must be related through the boundary conditions imposed on these infinitesimally thin world tubes.
\\The density that describes that point particle binary is given by
\begin{equation}
\rho=\frac{\delta(\theta-\pi/2)}{r^2}\left(M_1\delta(r-r_1)\delta(\phi-\tilde\phi)+ M_2\delta(r-r_2)\delta(\phi-\tilde\phi-\pi)\right),
\label{dens}
\end{equation}
where, $r_i\ (M_i)$ are the orbital radius (mass) of each particle, $r_1 < r_2$ and \ $\tilde\phi:=\tilde\phi(u)$ is the angular position as indicated in figure \ref{figure1}.\\
The instantaneous radius of the particles' orbits reads 
\begin{equation}
r_j=\frac{\mu d}{M_j}, \hspace{0.5cm} \mu=\frac{M_1M_2}{M_1+M_2}, \hspace{0.5cm} j=1,2,
\end{equation}
where the separation between the masses $d$ is given by
\begin{equation}
d=\frac{a(1-\epsilon^2)}{1+\epsilon \cos\tilde{\phi}},
\end{equation}
in which $\epsilon$ represents the eccentricity, and $a$ is a parameter which becomes the radius of the orbits when the eccentricity is zero. 
For Keplerian orbits, the angular velocity reads
\begin{equation}
\dot{\tilde{\phi}}=\frac{\sqrt{a(1-\epsilon^2)(M_1+M_2)}}{d^2}.
\end{equation} 
\\In order to reduce the field equations \eqref{field_eqs} to a system of ordinary differential equations with respect to the luminosity distance $r$, the metric functions are re-expressed as  linear combinations of the $\eth^s Z_{lm}$ spherical harmonics, 
\begin{equation}
_sf=\sum_{l,m} \Re(_sf_{lm} e^{i|m|\tilde\phi})\ \eth^s\  Z_{lm},
\label{sf}
\end{equation}
where $_sf$ represents the spin-weighted functions $\beta, w, U, J$, the symbol  $\sum_{l,m}$ indicates $\sum_{l=2}^\infty\sum_{m=-l}^l$, the coefficients $_sf_{lm}$ depend only on $r$ i.e., $_sf_{lm}:=\ _sf_{lm}(r)$, $Z_{lm}$ indicates $_0Z_{lm}$, and the angle $\tilde{\phi}$ is an arbitrary function of the retarded time $u$. This expansion directs our problem not to an
initial value, but instead to a boundary value problem.
\\
Thus, the functions \eqref{sf} are substituted in \eqref{field_eqs}, yielding
\begin{subequations}
\begin{align}
&\beta_{lm,r}= 2\pi \int_{\Omega}d\Omega\ \overline{Z}_{lm}\int_{0}^{2\pi}d\tilde\phi \ e^{-i|m|\tilde\phi} r T_{22},
\label{field_eq_s1}
\end{align}

\begin{align}
& -\frac{(l+2)(l-1)J_{lm,r}}{2} -\beta _{lm,r} +\frac{2 \beta_{lm}}{r} +\frac{\left(r^4 U_{lm,r}\right)_{,r}}{2r^2} \nonumber\\
&= \frac{8 \pi}{\sqrt{l(l+1)}} \int_{\Omega} d\Omega \ \overline{Z}_{lm}\int_{0}^{2\pi} d\tilde\phi \ e^{-i|m|\tilde{\phi}} T_{2A} q^A, 
\label{field_eq_s2}
\\[0.8cm]
& 2l(l+1) \beta_{lm} +(l-1)l(l+1)(l+2)J_{lm}  +\frac{l(l+1)\left(r^4\left(U_{lm}\right)\right)_{,r}}{r^2}\nonumber\\
& + 4 \beta_{lm} -2 w_{lm,r} = 8 \pi \int_{\Omega}d\Omega \ \overline{Z}_{lm}\int_{0}^{2\pi}d\tilde\phi\ e^{-i|m|\tilde{\phi}}\left(h^{AB} T_{AB}-r^2 T\right) , \label{field_eq_s3}\\[0.8cm]
&-2 \beta_{lm} + \left(r^2 U_{lm} \right)_{,r} - \left(r^2 J_{lm,r}\right)_{,r} +2 r\dot{\tilde{\phi}} \left(rJ_{lm}\right)_{,r} \nonumber\\
&= \frac{8 \pi}{\sqrt{(l-1)l(l+1)(l+2)}} \int_{\Omega}d\Omega \ \overline{Z}_{lm} \int_{0}^{2\pi}d\tilde{\phi}\ e^{-i|m|\tilde{\phi}}T_{AB}  q^A q^B,
\label{field_eq_s4}\\[0.8cm]
& -\frac{l(l+1) w_{lm}}{2 r^3} - \frac{l(l+1)\beta_{lm} }{r^2} +l(l+1)\dot{\tilde{\phi}} U_{lm}  +\frac{\dot{\tilde{\phi}} w_{lm}}{r^2} +\frac{w_{lm,rr}}{2 r} -\frac{2 \dot{\tilde{\phi}} \beta_{lm}}{r} \nonumber\\
&+\frac{2\beta_{lm,r}}{r} + \beta_{lm,rr}-2\dot{\tilde{\phi}}\beta_{lm,r}=8 \pi \int_{\Omega}d\Omega \ \overline{Z}_{lm} \int_{0}^{2\pi}d\tilde{\phi}\ e^{-i|m|\tilde{\phi}} \left(\frac{T}{2}+T_{11}\right), \label{field_eq_s5}
\\[0.8cm]
& -\frac{l(l+1)\beta_{lm} }{r^2} +\frac{l(l+1)\left(r^2 U_{lm} \right)_{,r}}{2r^2} +\frac{w_{lm,rr}}{2 r} \nonumber\\
&=8 \pi \int_{\Omega}d\Omega \ \overline{Z}_{lm} \int_{0}^{2\pi}d\tilde{\phi}\ e^{-i|m|\tilde{\phi}} \left(\frac{T}{2}+T_{12}\right), \label{field_eq_s6}\\[0.8cm]
& -\frac{(l+2)(l-1)J_{lm}\dot{\tilde{\phi}}}{2} +\frac{1}{2}\left(\frac{w_{lm}}{r}\right)_{,r} -\dot{\tilde{\phi}} \beta _{lm} +\frac{\left(r^4U_{lm,r}\right)_{,r}}{2r^2}\nonumber\\
& -\frac{r^2\dot{\tilde{\phi}}}{2}U_{lm,r} +U_{lm} =\frac{8 \pi}{\sqrt{l(l+1)}}  \int_{\Omega}d\Omega \ \overline{Z}_{lm} \int_{0}^{2\pi}d\tilde{\phi}\ e^{-i|m|\tilde{\phi}} T_{1A} q^A, \label{field_eq_s7}
\end{align}
\label{field_eq_s}%
\end{subequations}
where $d\Omega=d\phi \ d\theta \sin\theta$ is the solid angle element and the ortho-normality relations for the spin-weighted spherical harmonics $_sZ_{lm}$ is employed, i.e., 
\begin{equation*}
\int_{\Omega} d\Omega \ _sZ_{lm}\ _s\overline{Z}_{l'm'}=\delta_{ll'}\delta_{mm'}.
\end{equation*}
To integrate the above system of coupled ordinary differential equations, it is possible to show that equations \eqref{field_eq_s1}, \eqref{field_eq_s2} and \eqref{field_eq_s4} yield a fourth order differential equation for $J_{lm}$, which does not depend on any other metric variables. In order to do that, we re-express \eqref{field_eq_s1}, \eqref{field_eq_s2} and \eqref{field_eq_s4} as
\begin{subequations}
\begin{align}
&\beta_{lm,r}= A_{lm}, 
\label{field_eq_ss1}\\[0.6cm]
& -(l+2)(l-1)rJ_{lm,r} -2r\beta _{lm,r} +4 \beta_{lm} +4r^2 U_{lm,r}\nonumber\\
& +r^3 U_{lm,rr} = B_{lm}, 
\label{field_eq_ss2}\\[0.6cm]
& - r^2 J_{lm,rr}  +2 r\left(r\dot{\tilde{\phi}}  - 1 \right)J_{lm,r} +2r \dot{\tilde{\phi}} J_{lm} +2r U_{lm} + r^2 U_{lm,r}\nonumber\\
& -2 \beta_{lm}  = D_{lm},
\label{field_eq_ss4}
\end{align}
\label{field_eq_ss}%
\end{subequations}
where $A_{lm}:=A_{lm}(r)$, $B_{lm}:=B_{lm}(r)$ and $D_{lm}:=D_{lm}(r)$ and are explicitly defined as,
\begin{subequations}
\begin{align}
& A_{lm}=2\pi \int_{\Omega}d\Omega\ \overline{Z}_{lm}\int_{0}^{2\pi}d\tilde\phi \ e^{-i|m|\tilde\phi} r T_{22},\\
& B_{lm}=\frac{16 \pi r}{\sqrt{l(l+1)}} \int_{\Omega} d\Omega \ \overline{Z}_{lm}\int_{0}^{2\pi} d\tilde\phi \ e^{-i|m|\tilde{\phi}} T_{2A} q^A,\\
& D_{lm}= \frac{8 \pi}{\sqrt{(l-1)l(l+1)(l+2)}} \int_{\Omega}d\Omega \ \overline{Z}_{lm} \int_{0}^{2\pi}d\tilde{\phi}\ e^{-i|m|\tilde{\phi}}T_{AB}  q^A q^B.
\end{align}
\end{subequations}
Substituting $x=r^{-1}$ and after a few simple algebraic transformations, equation \eqref{field_eq_ss4} yields a second order differential equation for $\tilde{J}_{lm}$, i.e., 
\begin{align}
&  - 2x^4 \tilde{J}_{lm,xx} - 4x^2\left(2x  + \dot{\tilde{\phi}}\right)\tilde{J}_{lm,x} +2x\left(2  \dot{\tilde{\phi}} + x(l+2)(l-1) \right)\tilde{J}_{lm}  \nonumber\\
&   = x(x(2D_{lm}+ B_{lm}))_{,xx} - 2(x(2D_{lm}+ B_{lm}))_{,x} - x(x B_{lm,x})_{,x} \nonumber\\
& \hspace{0.5cm }+ 3xB_{lm,x},
\label{master_eq}
\end{align}
where $\tilde{J}_{lm}=J_{lm,xx}$, and $B_{lm}$ and $D_{lm}$ corresponding to source terms. It is important to notice that, for the vacuum case, \eqref{master_eq} becomes homogeneous and then coincides with the master equation presented in \cite{M13}(\cite{B05}) for general $l$ (for the case $M=0$ and $l=2$).
\\The master equation \eqref{master_eq} has analytical solutions for the vacuum. In this case, it does not depend explicitly on $m$. We solve it for each $l=2,3,\cdots$ and as expected the family of solutions for each $l$, namely $\tilde{J}_{lm}$, depends on two constants of integration. Then, integrating it two times we obtain explicitly the family of solutions $J_{lm}$, which depends on four constants of integration.  
\\Now, equations \eqref{field_eq_s} will be solved for the vacuum, i.e., for $T_{\mu\nu}=0$, for each $l=2,3,...$. In order to do this, note that from \eqref{field_eq_s1} $\beta_{lm}$ does not depend on $r$, therefore, these coefficients are constants along the radial coordinate $r$. Next, with the solutions for $J_{lm}$ and $\beta_{lm}$, the second order differential equation for $U_{lm}$, namely \eqref{field_eq_s2}, is solved analytically. After this, \eqref{field_eq_s3} is solved for $w_{lm}$ and with the remaining equations, the constraint equations, the system is fully solved, finally generating families of solutions that depend only on four constants of integration. It is important to note that \eqref{field_eq_s6} is immediately satisfied for the vacuum case, because $R_{12}=0$ is satisfied identically. Thus, the families of solutions that satisfy \eqref{field_eq_s} for the vacuum are, for example, for $l=2$ and $m\ne 0$ given by
\begin{subequations}
\begin{align}
\beta_{2m}(r)&=D_{1\beta 2m},
\label{gen_sol_1_l2}\\
J_{2m}(r)&=\frac{2 i D_{1\beta 2m}}{\dot{\tilde{\phi}}  r \left| m\right| }-\frac{D_{1J2m} (\dot{\tilde{\phi}}  r \left| m\right| -1) (\dot{\tilde{\phi}}  r \left| m\right| +1)}{6 r^3} \nonumber \\
& -\frac{i D_{2J2m} e^{2 i \dot{\tilde{\phi}}  r \left| m\right| } (\dot{\tilde{\phi}}  r \left| m\right| +i)^2}{8 \dot{\tilde{\phi}} ^5 r^3 \left| m\right| ^5} +\frac{D_{3J2m} (\dot{\tilde{\phi}}  r \left| m\right| -3 i)}{\dot{\tilde{\phi}}  r \left| m\right| }, \label{gen_sol_2_l2} \\
U_{2m}(r)&=\frac{2 D_{1\beta 2m} (\dot{\tilde{\phi}}  r \left| m\right| +2 i)}{\dot{\tilde{\phi}}  r^2 \left| m\right|
} -\frac{D_{1J2m} \left(2 \dot{\tilde{\phi}} ^2 r^2 \left| m\right|^2 + 4 i \dot{\tilde{\phi}}  r \left| m\right| +3\right)}{6
r^4} \nonumber\\
& -\frac{D_{2J2m} e^{2 i \dot{\tilde{\phi}}  r \left| m\right| } (2 \dot{\tilde{\phi}}  r \left| m\right| +3 i)}{8 \dot{\tilde{\phi}} ^5 r^4 \left| m\right|^5} -\frac{i D_{3J2m} \left(\dot{\tilde{\phi}} ^2 r^2 \left| m\right| ^2+6\right)}{\dot{\tilde{\phi}}  r^2 \left| m\right| } , 
\label{gen_sol_3_l2}
\end{align}
\begin{align}
w_{2m}(r)&= -10 r D_{1\beta 2m} +6 r D_{3J2m} (2+i \dot{\tilde{\phi}}  r \left| m\right| ) -\frac{3 i D_{2J2m} e^{2 i \dot{\tilde{\phi}}  r \left| m\right| }}{4 \dot{\tilde{\phi}} ^5 r^2 \left|m\right| ^5} \nonumber\\
&  -\frac{i D_{1J2m} ((1+i) \dot{\tilde{\phi}}  r \left| m\right| -i)	(1+(1+i) \dot{\tilde{\phi}}  r \left| m\right| )}{r^2}, \label{gen_sol_4_l2}
\end{align}
\label{gen_sol_l2}
\end{subequations}
\noindent where $D_{nFlm}$ are constants of integration, with $n$ labelling a particular constant and $F$ the metric function whose integration generate it. On the other hand, the solutions for $m=0$ and $l=2$, are given by
\begin{subequations}	
\begin{align}
\beta _{20}(r)&= D_{1\beta 20},
\label{gen_sol_1_l3}\\
J_{20}(r)&= \frac{1}{6} D_{2J20} r^2+\frac{2 D_{1\beta 20}}{3}+\frac{D_{4J20}}{r} + \frac{D_{1J20}}{6 r^3},
\label{gen_sol_2_l3}\\
U_{20}(r)&= -\frac{D_{1J20}}{2 r^4}+\frac{2 D_{1\beta 20}}{r}+\frac{r D_{2J20}}{3} + \frac{2 D_{4J20}}{r^2},
\label{gen_sol_3_l3}\\
w_{20}(r)&= -D_{2J20} r^3-2 D_{ 1\beta 20} r-\frac{D_{1J20}}{r^2}.
\label{gen_sol_4_l3}
\end{align}
\end{subequations}
Since the solutions for $m=0$ are not responsible for gravitational radiation, we then omit any mention of them from now.
\\As the metric functions must be regular at the interior of the world tubes to represent physical solutions, then the families of solutions for $r\in [0,r_1)$ are bounded. Expanding the functions in power series of $r$ around $r=0$, and setting the constants of integration such that those non-convergent terms becomes null, we find relationships between the constants of integration. This procedure reduces, for example, \eqref{gen_sol_l2} to a family of solutions that depends on one parameter, i.e.,
\begin{subequations}
\begin{align}
\beta_{2m-}(r)&= 0, \label{par_sol_int_1_l2}\\
J_{2m-}(r)&= \frac{D_{2J2m-}}{24 \dot{\tilde{\phi}}^5 r^3 \left| m\right|^5} \left(2 \dot{\tilde{\phi}}^3 r^3 \left| m\right|^3 - 3 i \dot{\tilde{\phi}}^2 r^2 \left| m\right|^2 e^{2 i \dot{\tilde{\phi}}  r \left| m\right| }   \right.\nonumber\\
&\left.-3 i \dot{\tilde{\phi}} ^2 r^2 \left| m\right|^2  + 6 \dot{\tilde{\phi}}  r \left| m\right|  e^{2 i \dot{\tilde{\phi}}  r \left| m\right| } +3 i e^{2 i \dot{\tilde{\phi}}  r \left| m\right| }-3 i\right),\label{par_sol_int_2_l2}\\
U_{2m-}(r)&= -\frac{i D_{2J2m-}}{24 \dot{\tilde{\phi}} ^5 r^4 \left| m\right| ^5}\left(2 \dot{\tilde{\phi}} ^4 r^4 \left| m\right|^4+6 \dot{\tilde{\phi}} ^2 r^2 \left| m\right| ^2   \right. \nonumber \\ 
& \left. - 6 i \dot{\tilde{\phi}}  r \left| m\right|  e^{2 i \dot{\tilde{\phi}}  r \left| m\right| } -12 i \dot{\tilde{\phi}}  r \left| m\right| +9 e^{2 i \dot{\tilde{\phi}}  r \left| m\right| }-9\right), \label{par_sol_int_3_l2}\\
w_{2m-}(r)&= \frac{D_{2J2m-}}{4 \dot{\tilde{\phi}}^5 r^2 \left| m\right|^5} \left(2 i \dot{\tilde{\phi}} ^4 r^4 \left| m\right|^4 + 4 \dot{\tilde{\phi}} ^3 r^3 \left| m\right| ^3 -6 i \dot{\tilde{\phi}} ^2 r^2 \left| m\right| ^2  \right. \nonumber \\
&\left. -6 \dot{\tilde{\phi}}  r \left| m\right| -3 i e^{2 i \dot{\tilde{\phi}}  r \left| m\right| }+3 i\right).  \label{par_sol_int_4_l2}
\end{align}
\end{subequations}
The families of solutions for $r\in (r_1,r_2)$ have the same structure such as \eqref{gen_sol_l2}, because there is not any physical restriction to impose on the solutions or to assure their convergence or yet to avoid any divergence. Whereas, for $r>r_2$ it is necessary to impose convergence at the null infinity. This is done by demanding that $D_{2Jlm+}=0$. As a result, one obtains a set of families of solutions that depends on eight parameters for the whole space-time. These parameters are $D_{2Jlm-}$, $D_{1\beta lm\pm}$, $D_{1Jlm\pm}$, $D_{2Jlm\pm}$, $D_{3Jlm\pm}$, $D_{1\beta lm+}$, $D_{1Jlm+}$, $D_{3Jlm+}$, where the subscript $-$, $\pm$ and $+$ indicates the zones $r<r_1$, $r_1<r<r_2$ and $r>r_2$, respectively. 
\\The presence of the binary in the space-time, induces jumps in the metric and in its derivatives just at the orbits, i.e.,
\begin{eqnarray}
& \left[w_{lm}(r_j)\right]=\Delta w_{jlm},& \hspace{0.2cm} \left[\beta_{lm}(r_j)\right]=\Delta \beta_{jlm}, \nonumber\\
& \left[J_{lm}(r_j)\right]=0, & \hspace{0.2cm} \left[U_{lm}(r_j)\right]=0,
\label{bound_cond_3}
\end{eqnarray}
and 
\begin{eqnarray}
& \left[w'_{lm}(r_j)\right]=\Delta w'_{jlm},& \hspace{0.2cm} \left[\beta_{lm}'(r_j)\right]=\Delta \beta_{jlm}', \nonumber\\
& \left[J'_{lm}(r_j)\right]=\Delta J'_{jlm}, & \hspace{0.2cm} \left[U'_{lm}(r_j)\right]=\Delta U'_{jlm},
\label{bound_cond_4}
\end{eqnarray}
where $j=1,2$, $|m|<l$, the brackets $[f_{lm}(r_j)]$ indicate 
\begin{equation}
[f_{lm}(r_1)]=f_{lm\pm}|_{r_1}-f_{lm-}|_{r_1},\hspace{0.5cm}\text{or} \hspace{0.5cm}[f_{lm}(r_2)]=f_{lm+}|_{r_2}-f_{lm\pm}|_{r_2}, 
\end{equation}
and $\Delta w_{jlm}$, $\Delta \beta_{jlm}$, $\Delta w_{jlm}'$, $\Delta \beta_{jlm}'$, $\Delta J'_{jlm}$ and $\Delta U'_{jlm}$ are functions to be determined. 
\\Solving equations \eqref{bound_cond_3} and \eqref{bound_cond_4}, one obtains
\begin{subequations}
\begin{align}
\Delta \beta_{jlm}&=b_{jlm},\label{bound_cond_5}\\
\Delta w_{jlm}&=-2r_jb_{jlm},\label{bound_cond_6}\\
\Delta J'_{jlm}&=\frac{8\dot{\tilde{\phi}} ^2 r_j  b_{jlm} \left|  m \right|^2} {(l-1)l(l+1)(l+2)}, \label{bound_cond_7}\\
\Delta U'_{jlm}&=2b_{ilm}\left(\frac{1}{r_i^2}-\frac{4i\dot{\tilde{\phi}} |m|}{l(l+1)r_i}\right), \label{bound_cond_8}
\end{align}	
\label{bound_conds}%
\end{subequations}
where $b_{jlm}$ are constants, which imply that $\Delta \beta'_{jlm}=0$. Also, the constants $D_{nFlm}$ depend on two parameters, namely $b_{1lm}$ and $b_{2lm}$. As an example, we show $D_{1J2m+}$ for $|m|\ne 0$, i.e.
\begin{align}
D_{1J2m+}=&\frac{i r_1^2 b_{12m} e^{-2 i r_1 \dot{\tilde{\phi }} \left| m\right|
}}{\dot{\tilde{\phi }} \left| m\right| }-\frac{i r_1^2 b_{12m}}{\dot{\tilde{\phi }} \left| m\right| }+\frac{2 r_1  b_{12m} e^{-2 i r_1\dot{\tilde{\phi }} \left| m\right| }}{\dot{\tilde{\phi }}^2 \left| m\right| ^2} +\frac{2 r_1 b_{\text{12m}}}{\dot{\tilde{\phi }}^2 \left| m\right|^2} \nonumber\\
&-\frac{3 i b_{12m} e^{-2 i r_1  \dot{\tilde{\phi }} \left| m\right| }}{\dot{\tilde{\phi }}^3
	\left| m\right| ^3} - \frac{3 b_{12m} e^{-2 i r_1  \dot{\tilde{\phi }} \left| m\right|
}}{r_1 \dot{\tilde{\phi }}^4 \left| m\right| ^4} -\frac{3 b_{12m}}{r_1 \dot{\tilde{\phi }}^4 \left| m\right| ^4}  -\frac{3i b_{12m}}{r_1^2 \dot{\tilde{\phi }}^5 \left| m\right| ^5}\nonumber\\
& +\frac{3 i b_{12m} e^{-2 i r_1 \dot{\tilde{\phi }} \left| m\right| }}{r_1^2 \dot{\tilde{\phi }}^5 \left| m\right| ^5} +\frac{3 i
b_{12m}}{\dot{\tilde{\phi }}^3 \left| m\right| ^3} +\frac{i r_2^2 b_{22m} e^{-2 i r_2  \dot{\tilde{\phi }} \left| m\right| }}{\dot{\tilde{\phi }}
\left| m\right| } -\frac{i r_2^2 b_{22m}}{\dot{\tilde{\phi }} \left| m\right| } \nonumber\\
& +\frac{2 r_2  b_{\text{22m}} e^{-2 i r_2  \dot{\tilde{\phi }} \left| m\right|
}}{\dot{\tilde{\phi }}^2 \left| m\right| ^2}+\frac{2 r_2  b_{22m}}{\dot{\tilde{\phi
}}^2 \left| m\right| ^2}-\frac{3 i b_{22m} e^{-2 i r_2  \dot{\tilde{\phi }} \left|
m\right| }}{\dot{\tilde{\phi }}^3 \left| m\right| ^3} -\frac{3b_{22m}}{r_2 \dot{\tilde{\phi }}^4 \left| m\right| ^4}\nonumber\\
& -\frac{3 b_{22m} e^{-2 i r_2 \dot{\tilde{\phi }} \left| m\right| }}{r_2 \dot{\tilde{\phi }}^4 \left| m\right| ^4}+\frac{3 i
b_{22m} e^{-2 i r_2  \dot{\tilde{\phi }} \left| m\right| }}{r_2^2 \dot{\tilde{\phi }}^5 \left| m\right| ^5}-\frac{3 i b_{22m}}{r_2^2\dot{\tilde{\phi }}^5 \left| m\right| ^5}+\frac{3 i b_{22m}}{\dot{\tilde{\phi }}^3\left| m\right| ^3}
\end{align}
The parameters $b_{jlm}$ for $j=1,2$ are  determined directly from \eqref{bound_cond_5} and \eqref{field_eq_s1}. In particular for the binary system, 
\begin{equation}
b_{jlm}=2M_j\int_0^{2\pi} d\tilde{\phi}\ \frac{e^{-i|m|\tilde\phi }\overline{Z}_{lm}(\pi/2,\tilde{\phi}+\pi \delta_{2j})}{r_j^2}.
\label{bes}
\end{equation}
where, it is important to note that the spin-weighted spherical harmonics $Z_{lm}$ become real on the equatorial plane $\theta=\pi/2$, but in general these functions are complex. 
\\Specifically, the non-null $b_{jlm}$, for the first $l$ and $m$, are given in Table 1.
\begin{table}[h!]
\begin{center}
\begin{tabular}{|c|c|c|c|c|}\hline
$l$ &  2 & 2&   3 &  3
\\\hline
$m$ & -2 & 0 &  -3 & -1  
\\\hline 
& & & &  \\[0.05cm]
$\dfrac{a \mu (\epsilon^2-1)b_{jlm}}{M_j^2}$ 
& $\dfrac{i\sqrt{15 \pi}}{2}$ & $\sqrt{5\pi}$  & $-\dfrac{i}{2}\sqrt{\dfrac{35 \pi}{2}}$ & $\dfrac{i}{2}\sqrt{\dfrac{21 \pi}{2}}$ 
\\[0.05cm]
& & & &  \\\hline
\end{tabular}
\end{center}
\caption{First non-null values for the constants $b_{jlm}$.}
\label{table1}
\end{table}
\\We write these coefficients only for $m<0$, because the others can be obtained remembering that,
\begin{equation}
_s\overline{Z}_{lm}=\left(-1\right)^{s+m}\ _{-s}Z_{l(-m)}.
\end{equation}
Thus, for $m\ne0$, 
\begin{equation}
b_{jlm}=ib_{jl(-m)}\hspace{0.5cm}j=1,2.
\end{equation}
\section{Gravitational radiation emitted by the binary}
The power emitted in gravitational waves is computed from the Bondi's News function, which in the linear regime of the Einstein's field equations \cite{B05} reads
\begin{equation}
\mathcal{N}=\lim\limits_{r\rightarrow\infty}\left(-\frac{r^2J_{,ur}}{2}+\frac{\eth^2\omega}{2}+\eth^2\beta\right).
\end{equation}
In terms of the coefficients $_sf_{lm}$, it reads
\begin{align}
\mathcal{N}=&\sum_{l,m}\lim\limits_{r\rightarrow\infty}\Re\left(\left(-\frac{ir^2\dot{\tilde{\phi}}|m| J_{lm,r}}{2}-\frac{r^2\dot{\tilde{\phi}} J_{lm,\tilde{\phi} r} }{2}\right.\right.\nonumber\\
&\hspace{2.5cm}+\left.\left.\frac{l(l+1)J_{lm}}{4}+\beta_{lm}\right)e^{i|m|\dot{\tilde{\phi}}}\right)\eth^2Z_{lm},
\label{news2}
\end{align}
where the sum indicates that the News is constructed from the contribution of several multipole terms. Here it is important to note that the coefficients $J_{lm}$ depend directly on the source angular position, represented by $\tilde{\phi}$, just as indicated in equation \eqref{master_eq}.  For this reason the time-retarded derivative $J_{,ur}$ is re-expressed using the chain rule.    
\\When the solutions to the field equations, for $r>r_2$ are substituted in \eqref{news2}, for $l=2$, one finds,
\begin{align}
\mathcal{N}=& 2i\sqrt{\frac{2}{3}} \dot{\tilde{\phi}} \left(\Re(e^{2i\tilde{\phi}}D_{2J22+} )\ _2Z_{2\ 2}+\Re(e^{2i\tilde{\phi}}D_{2J2-2+})\ _2Z_{2\ -2}\right)\nonumber\\
&+\frac{1}{2}\sqrt{\frac{3}{2}}\dot{\tilde{\phi}}\left(\Re(e^{2i\tilde{\phi}}D_{2J22+}')\ _2Z_{2\ 2}+\Re(e^{2i\tilde{\phi}}D_{2J2-2+}')\ _2Z_{2\ -2}\right),
\label{Nf}
\end{align}
where the prime indicates derivation with respect to $\tilde{\phi}$. It is worth noting that the $D_{2Jlm+}$ depend on $\dot{\tilde{\phi}}$, just as indicated in \eqref{master_eq}. Given that $\dot{\tilde{\phi}}:=\dot{\tilde{\phi}}(\tilde{\phi})$, then they are functions of the retarded angular position. Likewise, it is important to note that the absence of terms for $|m|=1$ in the News expression is because $b_{j21}=b_{j2-1}=0$ as indicated in Table \ref{table1}. In addition, despite $b_{jl0}\ne 0$ for $l=2,3,\cdots$, the terms for $m=0$ do not enter in the News, which indicates that they are non-radiative terms.
\\The power lost by gravitational radiation emission is computed by just integrating the square of the norm of the News in all directions, namely 
\begin{equation}
\frac{dE}{du}=\frac{1}{4\pi }\int_{\Omega}d\Omega\ |\mathcal{N}|^2.
\end{equation}
In the limit of low velocities, $r_1\dot{\tilde{\phi}}\ll c$, $r_2\dot{\tilde{\phi}} \ll c$ and for $l=2$, we find that the power reads
\begin{align}
\frac{dE}{du}=&\frac{32M_1^2M_2^2\left(M_1+M_2\right)\left(1+\epsilon\cos\tilde{\phi}\right)^6}{5a^5(1-\epsilon^2)^5}\nonumber\\
&+\frac{8 M_1^2M_2^2\left(M_1+M_2\right)\epsilon^2\sin^2\tilde{\phi}(1+\epsilon\cos\tilde{\phi})^4}{15 a^5(1-\epsilon^2)^5},
\label{power}
\end{align}
which is nothing but the Peters and Mathews expression for the energy lost by binary systems directly computed from the quadrupole radiation formulae  \cite{PM63}. 
\\
The agreement between our results and those by Peters and Mathews is in fact expected, since the system under study is the same. On the other hand, this agreement shows that the characteristic formalism in the linear regime has been properly applied in the present paper. Recall that Winicour in the 1980's decade showed that the Bondi's News function in the Quasi-Newtonian regime \cite{IWW85} is just $\mathcal{N}=\dddot{Q}$, with $Q=q^A\overline{q}^BQ_{AB}$ .
\\It is worth noting that, for the case of circular orbits, the two first terms in the News \eqref{Nf}, lead directly to
\begin{align}
\frac{dE}{du}=&\frac{32M_1^2M_2^2\left(M_1+M_2\right)}{5a^5}.
\end{align} 
Likewise, it is important to note that the first term in the power expression \eqref{power} represents approximately $97\%$ of the power emitted by the source for $\epsilon<0.5$. Thus, a reasonable approximation is just given by the first term of \eqref{power}.
\section{Summary and Conclusions}
We study for the first time in the literature a binary system composed of point particles of unequal masses in eccentric orbits in the linear regime of the characteristic formulation of general relativity. This work generalises previous studies \cite{B05} (\cite{CA15}) in which a system of equal (different) masses in circular orbits is considered.
	\\
We show that the boundary conditions on the time-like world tubes \eqref{bound_conds} can be extended beyond circular orbits. Concerning the power lost by the emission of gravitational waves, it is directly obtained from the Bondi's News function.
\\
Since the contribution of the several multipole terms ($l>2$) to the power is smaller than the contribution given by the mode $l=2$, the terms for $l>2$ are disregarded in the power expression \eqref{power} . In addition, the second term in \eqref{power} is small as compared to the first one. For example, for eccentricities $\epsilon \lesssim 0.5$ the first term contributes with almost $97\%$ of the power emitted in gravitational waves. 
\\
It is worth noting that our results are completely consistent, because we obtain the same result for the power derived by Peters and Mathews in a different approach. Recall that the News function in the Quasi-Newtonian limit corresponds to the third derivative with respect to the retarded time of the quadrupole moment contracted with the tangent vectors $q^A$, i.e., $\mathcal{N}=\dddot{Q}$, where $Q=q^A\overline{q}^BQ_{AB}$.
\\
In addition, the present study constitutes a powerful tool to construct extraction schemes in the characteristic formalism to obtain the gravitational radiation produced by binary systems during the inspiralling phase. This can be done in regions that are far enough from the sources where the space-time can be essentially considered flat.
 \\
Finally, it is worth mentioning that a proper and rigorous resolution of the problem here studied needs to be identified as an important outstanding matter for future research in this field. A step in that direction can be given, for example, by the introduction of a small parameter connecting the non-linear and the linear regimes, just as Winicour presents in his remarkable paper \cite{W83}. Such parameter would allow one to explore higher order perturbations as well as a rigorous solution of the problem considered in this paper.
\begin{acknowledgements}
We thank the Brazilian agencies CAPES, FAPESP (2013/11990-1) and CNPq (308983/2013-0) by the financial support. We would also like to thank the referees for their helpful comments and suggestions.
\end{acknowledgements}

\begin{thebibliography}{10}
\providecommand{\url}[1]{{#1}}
\providecommand{\urlprefix}{URL }
\expandafter\ifx\csname urlstyle\endcsname\relax
  \providecommand{\doi}[1]{DOI \discretionary{}{}{}#1}\else
  \providecommand{\doi}{DOI \discretionary{}{}{}\begingroup
  \urlstyle{rm}\Url}\fi

\bibitem{RBPS10}
C.~Reisswig, N.T. Bishop, D.~Pollney, B.~Szil\'agyi, Classical Quantum Gravity
  \textbf{27}(7), 075014 (2010)

\bibitem{BBSW11}
M.C. {Babiuc}, N.T. {Bishop}, B.~{Szil{\'a}gyi}, J.~{Winicour}, Phys. Rev. D
  \textbf{79}(8), 084011 (2009)

\bibitem{BSWS11}
M.C. {Babiuc}, B.~{Szil{\'a}gyi}, J.~{Winicour}, Y.~{Zlochower}, Phys. Rev. D
  \textbf{84}(4), 044057 (2011)

\bibitem{GPW94}
R.~G\'omez, P.~Papadopoulos, J.~Winicour, J. Math. Phys. \textbf{35}(8), 4184
  (1994)

\bibitem{BD96}
W.~Barreto, A.~Da~Silva, General Relativity and Gravitation \textbf{28}(6), 735
  (1996)

\bibitem{BD99}
W.~Barreto, A.D. Silva, Classical and Quantum Gravity \textbf{16}(6), 1783
  (1999)

\bibitem{BPR98}
W.~Barreto, C.~Peralta, L.~Rosales, Phys. Rev. D \textbf{59}, 024008 (1998)

\bibitem{BGLMW99}
N.T. Bishop, R.~G\'omez, L.~Lehner, M.~Maharaj, J.~Winicour, Phys. Rev. D
  \textbf{60}, 024005 (1999)

\bibitem{LFJMP01}
F.~{Linke}, J.A. {Font}, H.T. {Janka}, E.~{M{\"u}ller}, P.~{Papadopoulos},
  Astronomy and Astrophysics \textbf{376}, 568 (2001)

\bibitem{SFP01}
F.~Siebel, J.A. Font, P.~Papadopoulos, Phys. Rev. D \textbf{65}, 024021 (2001)

\bibitem{SFMP03}
F.~Siebel, J.A. Font, E.~M\"uller, P.~Papadopoulos, Phys. Rev. D \textbf{67},
  124018 (2003)

\bibitem{BGLMW05}
N.T. Bishop, R.~G\'omez, L.~Lehner, M.~Maharaj, J.~Winicour, Phys. Rev. D
  \textbf{72}, 024002 (2005)

\bibitem{B05}
N.T. Bishop, Classical Quantum Gravity \textbf{22}(12), 2393 (2005)

\bibitem{BPR11}
N.T. Bishop, D.~Pollney, C.~Reisswig, Classical Quantum Gravity
  \textbf{28}(15), 155019 (2011)

\bibitem{CA15}
C.E. {Cede{\~n}o M.}, J.C.N. {de Araujo}, ArXiv e-prints  (2015)

\bibitem{PM63}
P.C. Peters, J.~Mathews, Phys. Rev. \textbf{131}, 435 (1963)

\bibitem{BGLW96}
N.T. Bishop, R.~G\'omez, L.~Lehner, J.~Winicour, Phys. Rev. D \textbf{54}, 6153
  (1996)

\bibitem{BGLMW97}
N.T. Bishop, R.~G\'omez, L.~Lehner, M.~Maharaj, J.~Winicour, Phys. Rev. D
  \textbf{56}, 6298 (1997)

\bibitem{GPLPW97}
R.~G\'omez, L.~Lehner, P.~Papadopoulos, J.~Winicour, Classical Quantum Gravity
  \textbf{14}(4), 977 (1997)

\bibitem{BGHLW03}
N.T. Bishop, R.~G\'omez, S.~Husa, L.~Lehner, J.~Winicour, Phys. Rev. D
  \textbf{68}, 084015 (2003)

\bibitem{GBF07}
R.~G\'omez, W.~Barreto, S.~Frittelli, Phys. Rev. D \textbf{76}, 124029 (2007)

\bibitem{BBM62}
H.~Bondi, M.G.J. van~der Burg, A.W.K. Metzner, Proc. Phys. Soc. London Sect., A
  \textbf{269}(1336), 21 (1962)

\bibitem{S62}
R.K. Sachs, Proc. Phys. Soc. London Sect., A \textbf{270}(1340), 103 (1962)

\bibitem{RBLTS07}
C.~Reisswig, N.T. Bishop, C.W. Lai, J.~Thornburg, B.~Szilagyi, Classical
  Quantum Gravity \textbf{24}(12), S327 (2007)

\bibitem{HS15}
C.J. Handmer, B.~Szil\'agyi, Classical Quantum Gravity \textbf{32}(2), 025008
  (2015)

\bibitem{T07}
G.F. {Torres Del Castillo}, Revista Mexicana de Fisica Supplement
  \textbf{53}(2), 125 (2007)

\bibitem{NP66}
E.T. {Newman}, R.~{Penrose}, J. Math. Phys. \textbf{7}, 863 (1966)

\bibitem{GMNRS66}
J.N. {Goldberg}, A.J. {Macfarlane}, E.T. {Newman}, F.~{Rohrlich}, E.C.G.
  {Sudarshan}, J. Math. Phys. \textbf{8}, 2155 (1967)

\bibitem{ZGHLW03}
Y.~Zlochower, R.~G\'omez, S.~Husa, L.~Lehner, J.~Winicour, Phys. Rev. D
  \textbf{68}, 084014 (2003)

\bibitem{RBP13}
C.~Reisswig, N.~Bishop, D.~Pollney, Gen. Relativ. Gravit. \textbf{45}(5), 1069
  (2013)

\bibitem{Stephani_1990}
H.~{Stephani}, \emph{{General Relativity, An Introduction to the Theory of
  Gravitational Field}}, 2nd edn. (Cambridge University Press, Canada, 1990)

\bibitem{MTW_1973}
C.W. {Misner}, K.S. {Thorne}, J.A. {Wheeler}, \emph{{Gravitation}}
  (W.H.~Freeman and Co., San Francisco, USA, 1973)

\bibitem{M13}
T.~M\"adler, Phys. Rev. D \textbf{87}, 104016 (2013)

\bibitem{IWW85}
R.A. Isaacson, J.S. Welling, J.~Winicour, J. Math. Phys. \textbf{26}(11), 2859
  (1985)

\bibitem{W83}
J.~Winicour, J. Math. Phys. \textbf{24}(5), 1193 (1983)

\end{thebibliography}
\providecommand{\noopsort}[1]{}\providecommand{\singleletter}[1]{#1}%

\end{document}